\documentclass[journal,twocolumn]{IEEEtran}
\usepackage{amssymb}
\usepackage{amsfonts}

\usepackage{enumerate}

\usepackage{enumerate}
\usepackage{amsmath,amsthm}
\usepackage{mathtools}
\usepackage{algorithm}
\usepackage[noend]{algpseudocode}
\usepackage{float}
\usepackage{hyperref}
\usepackage{color}
\usepackage{makeidx}
\usepackage{bbm}
\usepackage{graphicx}
\usepackage{lipsum}
\usepackage{soul}
\usepackage{tabularx}
\usepackage{dsfont}
\usepackage{multirow}
\usepackage[table,xcdraw]{xcolor}

\usepackage{amsfonts}
\usepackage{times}
\usepackage{graphicx}
\usepackage{latexsym}
\usepackage{dsfont}
\usepackage{amssymb}
\usepackage{amsmath}
\usepackage{cite}
\usepackage{verbatim}
\usepackage{subfigure}

\def\bb0{{\mathbb{0}}}

\def\bb{{\mathbf{b}}}

\def\b0{{\mathbf{0}}}

\def\sf0{{\mathsf{0}}}

\usepackage{epstopdf}

\newcommand{\sref}[1]{{Section}~\ref{#1}}
\newcommand{\fref}[1]{{Fig.}~\ref{#1}}

\newcommand{\tref}[1]{{Table}~\ref{#1}}

\algnewcommand{\Initialize}[1]{%
	\State \textbf{Initialization:} \parbox[t]{.8\linewidth}{\raggedright #1}}

\newcommand{\magenta}[1]{\textcolor{black}{#1}}

\begin{document}
\title{Zone-Specific CSI Feedback for Massive MIMO: \\ A Situation-Aware Deep Learning Approach}
\author{Yu Zhang and Ahmed Alkhateeb \thanks{Yu Zhang and Ahmed Alkhateeb are with Arizona State University (Email: y.zhang, alkhateeb@asu.edu).}}
\maketitle

\begin{abstract}

Massive MIMO basestations, operating with frequency-division duplexing (FDD), require the users to feedback their channel state information (CSI) in order to design the precoding matrices.  Given the powerful capabilities of deep neural networks in learning quantization codebooks, utilizing these networks in compressing the channels and reducing the massive MIMO CSI feedback overhead has recently gained increased interest.  Learning one model, however, for the full cell or sector may not be optimal as the channel distribution could change significantly from one \textit{zone} (an area or region) to another. In this letter, we introduce the concept of \textit{zone-specific} CSI feedback. By partitioning the site space into multiple channel zones, the underlying channel distribution can be efficiently leveraged to reduce the CSI feedback. This concept leverages the implicit or explicit user position information to select the right zone-specific model and its parameters.  To facilitate the evaluation of associated overhead, we introduce two novel metrics named \textit{model parameters transmission rate} (MPTR) and \textit{model parameters update rate} (MPUR). They jointly provide important insights and guidance for the system design and deployment. Simulation results show that significant gains could be achieved by the proposed framework. For example, using the large-scale Boston downtown scenario of DeepMIMO, the proposed zone-specific CSI feedback approach can on average achieve around 6dB NMSE gain compared to the other solutions, while keeping the same model complexity.

\end{abstract}

\section{Introduction} \label{Intro}

Massive MIMO systems rely on accurate channel state information (CSI) to achieve the promising beamforming and multiplexing gains. 
In frequency division duplexing (FDD) systems, the downlink (DL) CSI needs to be first estimated at the user equipment (UE) and then sent back to the base station (BS).
The CSI feedback scheme adopted by the current 5G NR is not scalable  to systems with large numbers of antennas due to the excessive feedback overhead \cite{TS_38_214}.
To address this problem, deep learning based CSI feedback approach has been recently proposed and is deemed as a promising solution \cite{Wen2018}.
However, considering the potentially large channel variations in a given site, it is generally inefficient to utilize a single deep learning model with reasonable complexity to achieve satisfactory performance.
This motivates the development of new CSI compression approaches.

\begin{figure}[t]
	\centering
	\includegraphics[width=.9 \columnwidth]{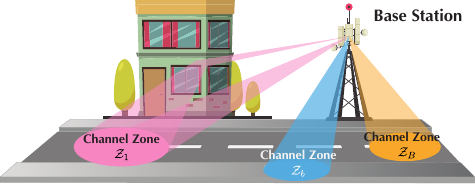}
	\caption{An illustration of the proposed zone-specific CSI feedback approach, where different deep learning weights are learned for each zone.}
	\label{fig:channel-zone}
\end{figure}

\textbf{Prior Work:}
Using deep learning for CSI feedback task is first demonstrated in \cite{Wen2018}, where a model named CsiNet shows clear advantages over some of the classical approaches.
Motivated by the promising performance, the following work extends the results by either using more complex/advanced deep learning models to improve the recovery accuracy \cite{Cui2022TransNet}, or by developing more effective model architectures and training schemes that are able to strike balance between complexity and performance \cite{Li2022CVLNet}.
However, the existing solutions directly process the channel samples from the whole site in a way that does not fully leverage the underlying channel distributions.
As a result, they normally require low compression rate (leading to high feedback overhead) or high model complexity in order to improve the CSI recovery accuracy.

\textbf{Contribution:}
In this letter, we propose a novel zone-specific CSI feedback framework that goes beyond the current site-specific framework.
Specifically, the site space is further divided into multiple channel zones, and by leveraging different CSI models at each zone, the channel samples can be more efficiently processed.
Having multiple CSI models also enables more flexible design/deployment, such as variable feedback length and model complexity. The selection of the zone leverages the situation awareness of the device/user either explicitly (using the position information) or implicitly (e.g., using CSI). 
Besides, we introduce two novel metrics named model parameter transmission and update rate that facilitates the evaluation of the overhead associated with the deep learning based CSI feedback approaches.
Simulation results highlight the potential gains of the proposed framework. 

\section{System and Channel Models} \label{sec:System}

We consider an FDD massive MIMO system where a BS with $N_t$ antennas is communicating with a single-antenna UE. We assume an OFDM based system with a total number of $K$ subcarriers. Therefore, in the downlink, the received signal at the $k$-th subcarrier is given by
\begin{equation}
  y_k = \overline{\mathbf{h}}_k^H\mathbf{f}_k x_k + z_k,
\end{equation}
where $\overline{\mathbf{h}}_k\in\mathbb{C}^{N_t\times1}$, $\mathbf{f}_k\in\mathbb{C}^{N_t\times1}$, $x_k\in\mathbb{C}$, and $z_k\in\mathbb{C}$ are the channel vector between the BS' antenna array and the UE's antenna, BS transmit beamforming, transmitted complex symbol, and the noise sample at the $k$-th subcarrier.
The transmit beamforming $\mathbf{f}_k$ at the BS requires the knowledge about the channel vector $\overline{\mathbf{h}}_k$. For that, the BS relies on the CSI feedback from the UE to determine $\mathbf{f}_k$.
For instance, in the current 5G NR \cite{TS_38_214}, CSI-RS transmitted by the BS is for the UE to estimate the DL channel, after which Type I/II codebooks are utilized by the UE to send the CSI back to the BS to schedule the DL data transmission.
For brevity and assuming perfect channel estimation, we denote the estimated channels across all the subcarriers as $\overline{\mathbf{H}}=[\overline{\mathbf{h}}_1,\dots,\overline{\mathbf{h}}_K]$.

We adopt a general geometric channel model for $\overline{\mathbf{h}}_k$ with $L_k$ paths \cite{Sayeed2002}. Each path $\ell$ has a complex gain $\alpha_k^{(\ell)}$ and an angle of departure (AoD) $\phi_k^{(\ell)}$, then the channel vector can be expressed as
\begin{equation}\label{eq:channel}
  \overline{\mathbf{h}}_k = \sum_{\ell=1}^{L_k}\alpha_k^{(\ell)}\mathbf{a}(\phi_k^{(\ell)}), ~ \forall k = 1,\dots,K,
\end{equation}
where $\mathbf{a}(\cdot)$ denotes the BS array response vector.

\section{Problem Formulation} \label{sec:Prob}

In this paper, we study the channel compression and recovery problem for the massive MIMO CSI feedback.
To reduce the feedback overhead, the original channel matrix $\overline{\mathbf{H}}$ is first transformed to the angular-delay domain using a 2D discrete Fourier transform (DFT) as given below \cite{Wen2018}
\begin{equation}\label{eq:DA}
  \mathbf{H} = \mathbf{F}_\mathrm{a}\overline{\mathbf{H}}\mathbf{F}_\mathrm{d}^H,
\end{equation}
where $\mathbf{F}_\mathrm{a}$ and $\mathbf{F}_\mathrm{d}$ are $N_t\times N_t$ and $K\times K$ DFT matrices, respectively.
After that, the UE compresses the transformed channel matrix $\mathbf{H}$ into a \emph{codeword} using a channel encoder
\begin{equation}
  \mathbf{s} = f_{\mathrm{en}}(\mathbf{H}),
\end{equation}
where $\mathbf{s}\in\mathbb{R}^{L}$ is a length-$L$ codeword.
The compression rate (CR), denoted as $\gamma$, achieved by the channel compression scheme is therefore defined as $\gamma = \frac{L}{2\times N_t \times N_c}$.
The codeword $\mathbf{s}$ is then reported to the BS through a feedback link.
Upon receiving such codeword, the BS decompresses the encoded information (i.e., $\mathbf{s}$) using a channel decoder to recover the original channel vectors, i.e., $\mathbf{H}$, which can be expressed as
\begin{equation}
  \widehat{\mathbf{H}} = f_{\mathrm{de}}(\mathbf{s}).
\end{equation}

Assuming that the channel encoder and decoder are parameterized deep learning models, the problem can then be cast as minimizing the channel recovery error for a given channel distribution under certain CR $\gamma$, formally
\begin{equation}
  \min_{\boldsymbol{\Theta}} ~ \mathbb{E}_\mathbf{H}\left[ \left\|\mathbf{H} - f_{\mathrm{de}}(f_{\mathrm{en}}(\mathbf{H};\boldsymbol{\Theta}_{\mathrm{en}});\boldsymbol{\Theta}_{\mathrm{de}})\right\|^2 \right],
\end{equation}
where $\boldsymbol{\Theta}=\left\{\boldsymbol{\Theta}_{\mathrm{en}},\boldsymbol{\Theta}_{\mathrm{de}}\right\}$
denotes the parameters of the model.

Despite the promising gain observed in prior deep learning based CSI feedback solutions \cite{Wen2018,Cui2022TransNet,Li2022CVLNet}, the existing methods do not fully leverage the underlying UE channel distribution.
\textbf{This could be an overlooked factor that has potential to further enhance the system performance.}
Specifically, different UE groups might experience different channel characteristics due to the distinct surrounding environments.
By explicitly leveraging such distribution information, it becomes possible to achieve higher channel recovery accuracy with similar or even less model complexity and feedback overhead.
This motivates the development of novel CSI feedback frameworks.

\section{Zone-Specific CSI Feedback}
In this section, we introduce our proposed zone-specific CSI feedback framework.

\subsection{Channel Zone}
For deep learning-based CSI feedback, a possible approach is for the UE to train one model per site; in what is defined as site-specific CSI feedback. In this paper, we propose to go beyond site-specific to what we call zone-specific CSI feedback. The motivation for this is to leverage the structure of the CSI distribution to further reduce the feedback overhead.
For instance, as illustrated in \fref{fig:channel-zone}, different sectors of the site tend to experience distinct channel distributions due to their own local propagation conditions.
\magenta{
Based on that, the channel model presented in \sref{sec:System} can be re-interpreted as a partitioning of the wireless environment into clusters of scatterers, where each one or more clusters represent a \textit{zone} from the BS's perspective, as illustrated in \fref{fig:channel-zone}, henceforth referred to as a \emph{channel zone}.
We denote such channel zone as $\mathcal{Z}_b$ with $b$ being the zone index.
It is worth noting that the channel zone approach can be viewed as a generalization of the site-specific concept, where the latter normally treats the whole cell as one zone.
}

Using Karhunen–Loève representation, one possible way for defining the zones is by expressing the channels of the users in each zone as
\begin{equation}\label{eq:cone_ch}
  \mathbf{h} = \mathbf V_{z_{b}} \boldsymbol{\Lambda}_{z_{b}}^{1/2} \mathbf{w}_{z_{b}}, ~ \forall \mathbf{h}\in \mathcal{Z}_b,
\end{equation}
where $\boldsymbol{\Lambda}_{z_{b}} \in \mathbb R^{r_b\times r_b}$ is a diagonal matrix with $r_b$ positive values occupying its diagonal. 
$\mathbf V_{z_{b}}\in \mathbb C^{KN_t\times r_b}$ is a sub-unitary matrix, i.e., $\mathbf V_{z_{b}}^H\mathbf V_{z_{b}}=\mathbf{I}_{r_b}$.
$\mathbf{w}_{z_{b}}\in \mathbb C^{r_b\times 1}$ is modeled as a random vector that is drawn from $\mathcal{CN}(\mathbf{0}, \mathbf{I}_{r_b})$.
{
Based on \eqref{eq:cone_ch}, the channels from the zone $\mathcal{Z}_b$ can be viewed as being sampled from a channel subspace spanned by $\mathbf V_{z_{b}} \boldsymbol{\Lambda}_{z_{b}}^{1/2}$. }

\subsection{Key Idea: Zone-Specific CSI Representation}

The motivation for partitioning a given site space into multiple zones comes from the observation that \textbf{the variation of the channel in a specific zone is much smaller than that in the whole site.}
The reduction in the channel variation makes it possible to achieve higher compression rate and/or better CSI recovery accuracy.
From machine learning perspective, this suggests that the compression can be realized on a manifold that has much lower dimensionality, which lays the foundations for the reduction in the CSI feedback overhead.

Based on that, we propose to decompose the single CSI feedback network into multiple (i.e., $B$) subnetworks with each one of them focusing on compressing and recovering the channels in one of the zones, that is
\begin{equation}
  f^{(b)}(\mathbf{h}_{z_b})=f_{\mathrm{de}}^{(b)}(f_{\mathrm{en}}^{(b)}(\mathbf{h}_{z_b};\boldsymbol{\Theta}_{\mathrm{en}}^{(b)});\boldsymbol{\Theta}_{\mathrm{de}}^{(b)}), ~\forall b,
\end{equation}
where $\mathbf{h}_{z_b}$ is to indicate that the channel sample is from the channel zone $\mathcal{Z}_b$.
Moreover, for simplicity, we assume that all these subnetworks have the same model architecture where only their parameters differ.
Specifically, for the $b$-th subnetwork, we denote its parameter as $\boldsymbol{\Theta}^{(b)}=\{\boldsymbol{\Theta}_{\mathrm{en}}^{(b)}, \boldsymbol{\Theta}_{\mathrm{de}}^{(b)}\}$. All the these subnetworks' parameters constitute the parameters of the composite model, i.e., $\boldsymbol{\Theta}=\{\boldsymbol{\Theta}^{(1)},\dots,\boldsymbol{\Theta}^{(B)}\}$.
The problem can then be cast as
\begin{equation}
  \min_{\{\boldsymbol{\Theta}^{(b)}\}_{b=1}^B} \mathbb{E}_{\mathbf{h}}\left[\min_{b}\|f^{(b)}(\mathbf{h};\boldsymbol{\Theta}^{(b)})-\mathbf{h}\|^2\right].
\end{equation}
For simplicity, we assume that each subnetwork, $f^{(b)}$, is independently trained with a given dataset, as will be discussed in the next subsection.

\subsection{Model Training}
The model is trained by leveraging a downlink channel dataset, denoted as $\boldsymbol{\mathcal{H}}=\{\mathbf{h}_1, \dots, \mathbf{h}_U\}$, which is collected by the system.
By applying channel clustering\footnote{Here, it is not restricted to any specific channel clustering method. In fact, the development of efficient channel clustering algorithm is crucial in reducing the CSI feedback overhead.}, the original channel dataset can be partitioned into $B$ non-overlapping channel subsets, expressed as
\begin{equation}\label{UE-clus}
  \boldsymbol{\mathcal{H}} = \boldsymbol{\mathcal{H}}^{(1)}\cup\cdots\cup\boldsymbol{\mathcal{H}}^{(B)},
\end{equation}
where $\boldsymbol{\mathcal{H}}^{(b)}\cap\boldsymbol{\mathcal{H}}^{(b^\prime)} = \emptyset, ~ \forall b\ne b^\prime ~\mathrm{and}~ \forall b, b^\prime\in\{1,\dots,B\}$.
Correspondingly, there are $B$ subnetworks with each of them trained on one channel subset.
We follow an end-to-end learning approach and the mean squared error (MSE) loss function is used for training the model
\begin{equation}
  \mathcal{L}_{\boldsymbol{\Theta}^{(b)}} = \frac{1}{|\boldsymbol{\mathcal{H}}^{(b)}|}\sum_{\mathbf{h}\in\boldsymbol{\mathcal{H}}^{(b)}}\left\|\mathbf{h}-f^{(b)}(\mathbf{h};\boldsymbol{\Theta}^{(b)})\right\|^2, ~\forall b.
\end{equation}
The $B$ trained subnetworks collectively constitute the composite CSI feedback model.

\section{Model Parameters \\ Transmission and Update Rates}\label{sec:MPTR}

In this section, we introduce two novel performance metrics, namely model parameters transmission rate (MPTR) and model parameters update rate (MPUR).
These metrics serve as key performance indicators to measure the overheads of different deep learning-based CSI feedback solutions.
The understanding/evaluation of these overheads becomes increasingly important, as each site will have multiple CSI encoders that the UE devices need to load and use.
This is different from what is currently adopted in the 5G NR system, where the compression codebook (e.g. Type I/II codebooks, etc.) is known before the actual deployment \cite{TS_38_214}.

\subsection{Spatial Zone and UE Mobility} \label{subsec:SZUM}

To define MPTR and MPUR, we first introduce the concept of \emph{spatial zone}. A spatial zone $\mathcal{S}_b$ is the counterpart of the channel zone $\mathcal{Z}_b$ in the spatial domain,
and it can be defined via finding a set of positions where their corresponding channels are all within the same channel zone, that is
\begin{equation}\label{spatial_zone}
  \mathcal{S}_b = \left\{\mathbf{x}\left|\mathbf{g}\left(\mathbf{x}\right)=\mathbf{h}, \forall \mathbf{h}\in\mathcal{Z}_b\right.\right\}, ~ \forall b\in\{1,\dots,B\},
\end{equation}
where $\mathbf{g}(\cdot)$ denotes the mapping function from position to channel.
In other words, \eqref{spatial_zone} implies that the whole cell space, denoted as $\mathcal{S}$, can be partitioned into non-overlapping spatial zones, that is, $\mathcal{S}=\bigcup_{b=1}^B\mathcal{S}_b$.

For a CSI feedback method that includes multiple channel encoders, the additional overhead is also closely related to the UE mobility pattern which normally triggers the model update.
Therefore, we assume that for any given UE, its mobility pattern is a realization of a random process $\mathcal{M}(t)$, and different UEs' mobility patterns are independent realizations of $\mathcal{M}(t)$.
By leveraging a discretized time series $\mathcal{T}=\{t_1,\dots,t_P\}$, where $t_1<\cdots<t_P$, we define the number of times of (spatial) zone switching within a certain time horizon (i.e., $t_P-t_1$) as
\begin{equation}
  N_{\mathrm{zs}} = \mathbb{E}_\mathcal{M}\left[\sum_{p=1}^{P-1}\mathbbm{1}\left\{\mathcal{M}(t_p)\in\mathcal{S}_b~\mathrm{and}~\mathcal{M}(t_{p+1})\in\mathcal{S}_b^\prime\right\}\right],
\end{equation}
where $\mathbbm{1}(\cdot)$ is the indicator function and $b\ne b^\prime$.
The rate of the zone switching can then be defined as $r_{\mathrm{zs}}=\frac{N_{\mathrm{zs}}}{t_P-t_1}$.

\subsection{MPTR and MPUR} \label{subsec:MPTR}

With the help of the above definitions of spatial zone and UE mobility, we can now define MPTR and MPUR as follows:
\begin{itemize}
  \item \textbf{MPTR:} The average rate with which the model parameters are downloaded in the UE device. It has an unit of number of parameters per second.
  \item \textbf{MPUR:} The frequency with which a UE device updates/switches its CSI encoder due to mobility.
\end{itemize}
MPTR, on one hand, allows the system to analyze the additional over-the-air overhead, in addition to the CSI feedback overhead, of a proposed deep learning-based CSI acquisition solution.
MPUR, on the other hand, characterizes the local behavior/overhead of a solution inside the UE device.
These metrics provide useful design considerations and evaluate the practicality of a solution under given scenarios.

It is important to note that model update is, in general, different from model download. The former is to update the encoder parameters, while the latter is to download a CSI encoder from BS into the UE device, which incurs the actual over-the-air overhead. Furthermore, \textbf{the model update is also normally more frequent than model download.}
In this paper, we assume that the model update is triggered by the zone switching. Therefore, the MPUR, denoted as $r_{\mathrm{mu}}$, is the same as zone switching rate, i.e., $r_{\mathrm{mu}}=r_{\mathrm{zs}}$, while the model downloading rate, denoted as $r_{\mathrm{md}}$, satisfies $r_{\mathrm{md}}\le r_{\mathrm{mu}}$.
For instance, if the UE device can download all the subnetworks at once, there will be no model downloading required when the UE changes its zone.
Based on the above definitions, assume the CSI encoder in each subnetwork has $V$ parameters, then, the MPTR of a specific method is equal to $Vr_{\mathrm{md}}$.
It can be inferred that the ``worst'' case scenario is when $r_{\mathrm{md}} = r_{\mathrm{mu}}$, i.e., the UE device can only store one CSI encoder at a time. Then, the MPTR of the given scheme becomes $Vr_{\mathrm{mu}}$.

\begin{figure}[t]
	\centering
	\includegraphics[width=.7\columnwidth]{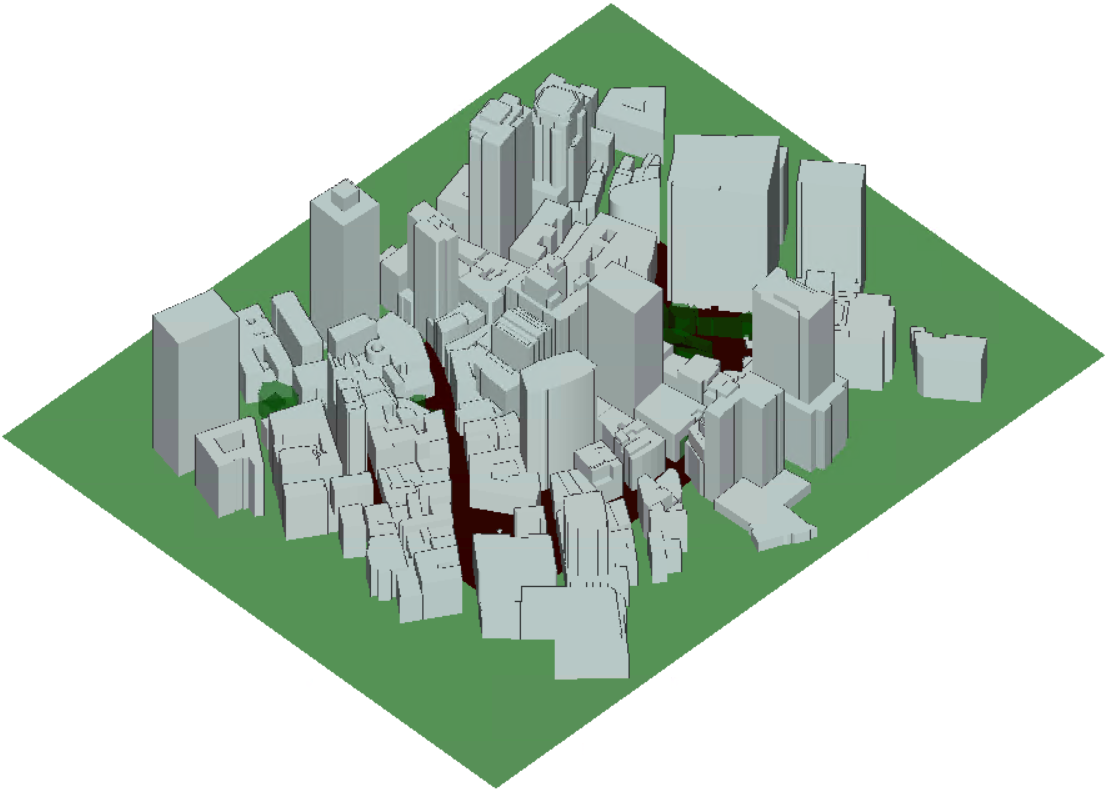}
    \caption{The 3D ray-tracing model of the adopted Boston downtown scenario.}
	\label{fig:scenario}
\end{figure}

\begin{figure}[t]
	\centering
	\includegraphics[width=.75\columnwidth]{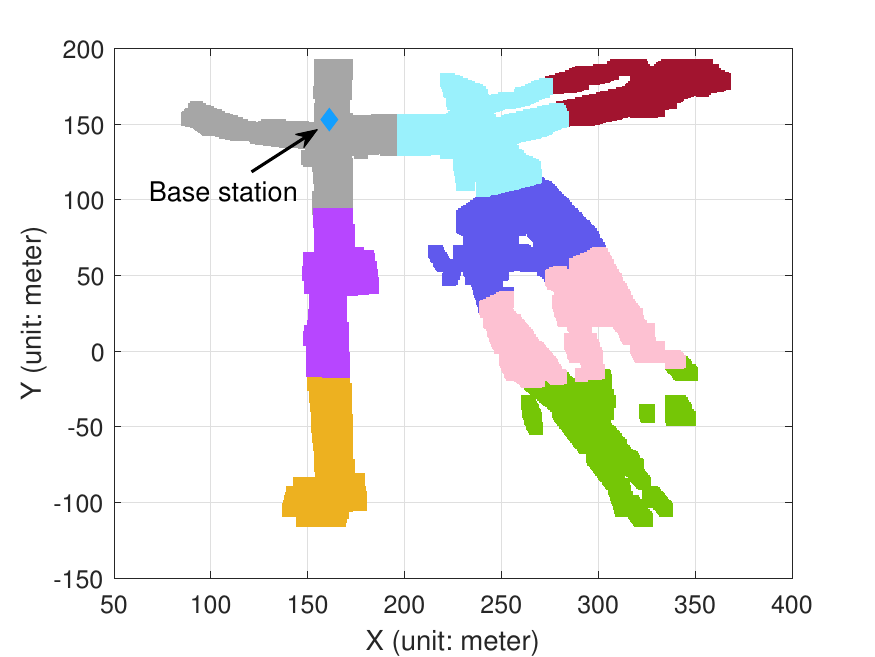}
    \caption{The spatial zones in the considered outdoor scenario.}
	\label{fig:zone}
\end{figure}

\section{System Operation}

In this section, we provide a few remarks regarding the operation of the proposed zone-specific CSI feedback approach.
One important question that needs to be answered is how to cluster the channel into channel zones.
In this work, we assume that UE position information is available to the network or to the cloud where the channel clustering will be performed.
Therefore, a natural idea is to partition the UEs based on their positions.
Intuitively, the spatial proximity also implies correlations in the channels, which is normally the case when there exists dominant multi-path components.
Moreover, partitioning UEs based on positions can effectively reduce the channel zone switching rate, which leads to a reduced MPTR/MPUR.
Given the availability of UE positions, we define an \emph{augmented} channel dataset that includes such information, that is, $\boldsymbol{\mathcal{H}}_{\mathrm{aug}}=\{(\mathbf{h}_1, \mathbf{x}_1), \dots, (\mathbf{h}_U, \mathbf{x}_U)\}$, where each sample consists of the UE channel and its position.

\noindent\textbf{Training phase:} The training process includes UE clustering and network training. In this case, we assume that the clustering is performed based on position data.
As a result, the set $\mathcal{S}$ of the user positions in the cell is partitioned into $B$ spatial zones.
Based on these  zones, the channels can be partitioned correspondingly.
After that, $B$ subnetworks will be trained based on the different channel subsets.
The training results include a \emph{position classifier} and a collection of subnetworks.

\noindent\textbf{Deployment phase:}
During the actual operation, both sides can select the subnetwork that corresponds to the current spatial zone based on the position information (i.e., using the trained position classifier to get the spatial zone information).

\textbf{Remark:}
It is important to notice that from reducing MPTR/MPUR perspective, it is more favorable to adopt position-based UE clustering method, which normally incurs less zone switching. This, however, might not be optimal from CSI feedback perspective (compression rate and CSI reconstruction accuracy). This motivates future research for developing channel clustering solutions that compromise between the MPTR/MPUR and CSI feedback reduction.

\section{Simulation Results} \label{sec:Simu}

\begin{figure}[t]
	\centering
	\includegraphics[width=.83\columnwidth]{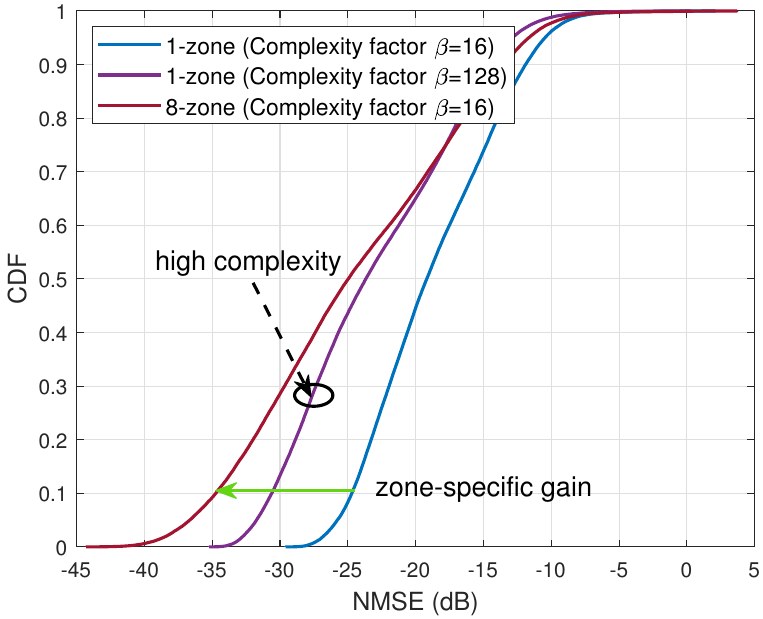}
	\caption{The NMSE performance with three different methods that have different neural network complexities, all with a compression rate of $1/64$.}
	\label{fig:nmse}
\end{figure}

\begin{table}[t]
	\caption{CSI feedback model parameters}
	\centering
	\resizebox{0.8\columnwidth}{!}{%
		\begin{tabular}{c|c|c|c}
			\hline
			\hline
			\multicolumn{2}{c|}{\textbf{Layer}} & \textbf{Dimension} & \textbf{Parameters} \\
			\hline
			\multirow{3}{*}{\textbf{Encoder}} &
			Fully-connected & $(2N_tN_c, \beta L)$ & $(2N_tN_c+1)\beta L$ \\
			& BatchNorm & $(\beta L,)$ & $2\beta L$ \\
			& Fully-connected & $(\beta L, L)$ & $(\beta L+1)L$ \\
			\hline
			\multirow{3}{*}{\textbf{Decoder}} &
			Fully-connected & $(L, \beta L)$ & $(L+1)\beta L$ \\
			& BatchNorm & $(\beta L,)$ & $2\beta L$ \\
			& Fully-connected & $(\beta L, 2N_tN_c)$ & $(\beta L+1)2N_tN_c$ \\
			\hline
			\hline
		\end{tabular}%
	}
	\label{tab:csiencoder}
\end{table}

\begin{table*}[t]
	\caption{The comprehensive performance evaluation of the three different feedback approaches}
	\centering
	\begin{tabular}{c|c|c|c|c|c}
		\hline
		\hline
		\textbf{Method} & \textbf{Mean NMSE} & \textbf{MPTR} & \textbf{MPUR} & \textbf{Multiplications} & \textbf{Parameters} \\
		\hline
		1-zone ($\beta=16$) & $-18.6490$ dB & $1184.16$ param/s & 0 & 4,261,888 & 4,262,976 / 8,529,984 \\
		1-zone ($\beta=128$) & $-22.9270$ dB & $9473.15$ param/s & 0 & 34,095,104 & 34,103,360 / 68,210,752 \\
		8-zone ($\beta=16$) & $-24.3731$ dB & $9473.28$ param/s & $0.0147$/s & 4,261,888 & 34,103,808 / 68,239,872 \\
		\hline
		\hline
	\end{tabular}
	\label{tab:compare}
\end{table*}

To evaluate the proposed solution, we leverage the DeepMIMO Boston5G scenario \cite{DeepMIMO}, where a BS is serving UEs in a downtown sector of Boston, as shown in \fref{fig:scenario}. 
The BS adopts a $64$-element (with $16$-by-$4$ panel configuration) uniform planar array (UPA) operating at a carrier frequency of $3.5$ GHz.
The dominant $15$ multi-paths of each BS-UE channel are considered.
Based on these configurations, a total number of $105,996$ UE channels are generated, out of which $24,000$ samples are used for training the model and the remaining samples are for testing the model.
We adopt a fully-connected layer based auto-encoder architecture as the CSI encoder and decoder networks\footnote{It is worth noting that the proposed zone-specific CSI feedback framework is not restricted to a specific deep learning architecture.}. The details of the model architecture can be found in \tref{tab:csiencoder}, where $\beta$ is a scaling factor that can change the model size.

\subsection{Numerical Results}
We first study the CSI recovery accuracy with and without multiple channel zones, where these zones are generated based on the UE positions.
We plot the empirical cumulative distribution function (CDF) of the achieved channel NMSE accuracy in \fref{fig:nmse}.
We compare the performance of three different models, where two of them have only single CSI feedback network for the entire site, with two model sizes (i.e., $\beta=16$ and $\beta=128$).
The third model has $8$ CSI feedback networks ($8$ zones) with each network the same size of the smaller single model, i.e., $\beta=16$. This makes the total number of parameters of the $8$ networks roughly the same as that of the single network with $\beta=128$.
Moreover, to further ensure the fairness between different experiments, we make the total number of training samples the same regardless of the number of partitioned  zones and different model sizes.
The compression rate for all the experiments is  $1/64$.

\textbf{Performance enhancement:}
\fref{fig:nmse} shows that by leveraging channel zones, the proposed zone-specific CSI feedback solution brings noticeable gains over the other two solutions.
\textbf{For instance, with the same model complexity, around $50\%$ of the UEs have less than $-25$ dB NMSE when there are $8$ channel zones, in contrast with only $8\%$ of UEs in the single zone case.}
Further, for the same scenario, $50\%$ of the users achieve more than $7$ dB gain in their CSI NMSE performance.
When comparing with the single-zone model that has $8$ times higher complexity ($\beta=128$), the proposed solution is still able to achieve better performance.
This clearly highlights the advantage of the zone-specific CSI feedback framework. 

\textbf{Comprehensive evaluation:}
The performance enhancement brought by having more zones, however, does not come free.
For instance, the UEs are required to download a collection of CSI encoders from the BS or the cloud, and they need to frequently update the CSI encoder parameters whenever a model update is triggered.
As a result, having more zones generally results in an increased MPTR. 
It is worth mentioning here that a high MPUR could also incur extra latency as well as increased UE power consumption.

Therefore, to comprehensively understand the system performance and cost, we also investigate the MPTR/MPUR and complexity of the different models.
To study the MPTR, considering that only the CSI encoders need to be distributed to the UE devices, the trained network parameters of the encoder constitute the major part of the over-the-air model transmission overhead.
Therefore, only the CSI encoder network architecture in \tref{tab:csiencoder} needs to be considered.
%
%
In addition to the model size, as discussed in \sref{sec:MPTR}, the MPTR/MPUR also depends on the model update rate, i.e., $r_{\mathrm{mu}}$,
%
%
which is mainly determined by the user mobility pattern.
%
To estimate $r_{\mathrm{mu}}$, in the simulation, we observe a time horizon of $T$ seconds with UE moving randomly within the cell space $\mathcal{S}$ under certain mobility, by which we count the number of model updates, denoted as $\widehat{N}_{\mathrm{mu}}$.
The model update rate can then be estimated as $\widehat{r}_{\mathrm{mu}}=\widehat{N}_{\mathrm{mu}}/T$, or equivalently, $T_{\mathrm{mu}}=T/\widehat{N}_{\mathrm{mu}}$ is the average time duration that the UE is using the same CSI encoder.

In \tref{tab:compare}, we summarize the comparison results of the three models shown in \fref{fig:nmse}. We consider a scenario where a pedestrian carrying a mobile device randomly walking around the region. A $10$ km/h mobility is assumed and a time window of $T=3,600$ seconds is simulated.
From the results, it can be seen that when comparing the $8$-zone solution with the small single-zone solution, it achieves $5.7$ dB improvement of the mean NMSE performance with a cost of larger overall model parameters (i.e., $8$ times of the other one) and large MPTR/MPUR.
When comparing the $8$-zone solution with the large single-zone solution, they have almost the same total number of parameters and MPTR, while the $8$-zone solution is still able to achieve better NMSE performance.
However, special attention should be paid on the computation complexity where \textbf{the number of multiplications that needs to be done in UE for each CSI feedback of the large single-zone solution is $8$ times of that of the $8$-zone solution.}
Given the limited computation and power capability of the UE device, such large amount of computations will likely introduce additional feedback latency and drastically shorten the battery life in the single-zone scenario. This gives another advantage to the proposed zone-specific CSI feedback approach.

\section{Conclusions and Future Work}

In this letter, we propose a novel zone-specific CSI feedback framework that brings noticeable gain over the site-specific approaches.
By training the models on channel zones with reduced variations, the composite CSI feedback framework can more efficiently leverage the underlying channel distribution.
We also introduce two novel metrics coined model parameter transmission and update rate to characterize the new overhead associated with the deep learning based CSI feedback approaches, providing important insights for future system design and deployment.
Important future extensions of this work include developing joint UE clustering and MPTR/MPUR optimization framework, intelligent model update methods, and zone-specific network design and dataset utilization.

\bibliographystyle{IEEEtran}

\end{document}